\documentclass{jpsj-suppl}
\usepackage{amsmath}
\usepackage{amssymb}
\usepackage{bm}
\usepackage{braket}
\usepackage{graphicx,epstopdf}

\title{Matrix-Product Ansatz for Excited States of Fractional Quantum
Hall Systems}

\author{Zheng-Yuan \textsc{Wang} and Masaaki \textsc{Nakamura}}
\inst{Department of Physics, Tokyo Institute of Technology, O-Okayama,\\
Meguro-ku, Tokyo 152-8551, Japan}

\email{wang@stat.phys.titech.ac.jp, masaaki@stat.phys.titech.ac.jp}

\recdate{January 24, 2013}

\abst{
Recently, it was discussed that the $\nu=1/3$ fractional quantum Hall
state can be expressed by a one-dimensional lattice model with an exact
matrix-prduct ground state, when toroidal boundary conditions are
assumed for a narrow strip [Phys. Rev. Lett. {\bf 109} (2012) 016401].
In this article, we discuss how the excitation spectra of this system
are calculated analytically based on the matrix-product formalism.  We
introduce variational wave functions for various momenta and optimize
them.  The obtained spectra of the charge neutral excitations show the
magneto-roton behavior.}
\kword{fractional quantum Hall effect, thin-torus limit, matrix product
state, exactly solvable model, excited states, magneto-roton}

\begin{document}
\maketitle

\section{Introduction}
In recent years, there have been theoretical efforts to study the
fractional quantum Hall (FQH) states around the thin-torus (or
Tao-Thouless, TT) limit, where the original 2D system in a magnetic
field with toroidal boundary conditions is reduced to 1D lattice models
with short-range interactions
\cite{Tao-T,Rezayi-H,Bergholtz-K,Seidel-F-L-L-M}.  This approach is
justified because the FQH states in this anisotropic system are known to
be adiabatically connected to the isotropic cases \cite{Rezayi-H}.  From
this point of view, the authors with Bergholtz have shown that the
$\nu=1/3$ Laughlin state \cite{Laughlin} can be expressed by a 1D model
with an exact matrix-product (MP) ground state \cite{Nakamura-W-B}.  In
this article, we give the details to obtain the excitation spectra of
this model based on the MP wave function with variational parameters
which partially mentioned in Ref.~\cite{Nakamura-W-B}, and generalize
this calculation to arbitrally momenta and obtain roton behavior of the
excitation spectra \cite{Girvin-M-P}.

\section{Fractional quantum Hall states on torus geometries}
We consider a model of $N_e$ interacting electrons in a magnetic field
$B$ on a torus boundary conditions with circumferences $L_1$ and $L_2$
for $x$ and $y$-directions, respectively.  For simplicity, we normalize
$x, y, L_1, L_2$ by magnetic length $l_B=\sqrt{c\hbar/eB}$.  When the
torus is pierced by $N_s$ magnetic flux quanta, boundary conditions
require the relation $L_1 L_2=2\pi N_s$.  In Landau gauge
($\bm{A}=By\widehat{x}$), a complete basis of $N_s$ degenerate
single-particle states in the lowest Landau level can be chosen as
\begin{equation}
\label{WavFunTLL}
\psi_i(x,y)=\frac{1}{\sqrt{\pi^{1/2} L_1}}
\sum_{j \in \mathbb{Z}}
\exp \left[ \mathrm{i}
\left(\frac{2\pi}{L_1}i+L_2j \right) x \right]
\exp \left[ -\frac{1}{2}
\left(y+\frac{2\pi }{L_1}i+L_2j \right)^2 \right],
\end{equation}
where the guiding center of the Gaussian function in $y$-direction is
given by the quantized momentum in $x$-direction with $i=0, 1, \cdots,
N_s-1$.  In this basis, any translationally-invariant 2D Hamiltonian
with two-body interactions assumes the following 1D lattice model,
\begin{equation}
\label{1D_model}
\mathcal{H}=\sum_{|m|< k \le N_s/2} \hat{V}_{km},
\quad
\hat{V}_{km}
\equiv \sum_{i} V_{km}^{\mathstrut}
c^\dagger_{i+m} c^\dagger_{i+k}
c_{i+k+m}^{\mathstrut}c_{i}^{\mathstrut},
\end{equation}
where the matrix-element $V_{km}$ specifies the amplitude for a pair
hopping process where two particles separated $k+m$ sites hop $m$ steps
to opposite directions (note that $m$ can be negative).  The $m=0$ terms
are the electrostatic repulsions, the matrix elements are real and
antisymmetric functions for interchange of $k$ and $m$.  This model
conserves the center of mass coordinate defined by $K=\sum_{i=1}^{N_s} i
n_i \ (\mathrm{mod} \ N_s)$ with $n\equiv
c^\dagger_{i}c_{i}^{\mathstrut}$, which is identified as the total
momentum in $x$-direction [see Eq.~(\ref{WavFunTLL})].

For small $L_1$ the matrix elements $V_{km}$ are simplified
considerably: Since one finds that for generic interaction,
the $m \neq 0$ terms are exponentially suppressed as
\begin{equation}
 \label{LeaShort}
  V_{km} \sim V_{k0}\exp \left(-2\pi^2 m^2/L_1^2 \right).
\end{equation}
Therefore, in the TT limit $L_1 \rightarrow 0$, $m \neq 0$ terms in the
original Hamiltonian (\ref{1D_model}) can be neglected, so that the
Hamiltonian can be rewritten as $\mathcal{H}_{\rm TT}=\sum_i \sum_j
V_{k0}n_{i+j} n_{j}$. Then the ground state is obtained at filling
factor $\nu=1/3$ is gapped, and three-fold degenerate charge density
wave state with three sites periodicity,
$\ket{\Psi_0}=\ket{010|010|010\cdots}$. A system of thin but finite
torus can be considered by taking the leading hopping terms
$\hat{V}_{km}$ $(m\neq 0)$ into account for the TT state $\ket{\Psi_0}$.

\section{Matrix-product ground state for $\nu=1/3$ Laughlin state}

The exactly solvable model for $\nu=1/3$ states \cite{Nakamura-W-B} can
be obtained by truncating the Hamiltonian (\ref{1D_model}) up to the
third nearest interactions $(k+|m| \le 3)$ as an approximation in the
vicinity of the TT limit.  When we further assume $L_2\to\infty$ limit,
the matrix elements of the pseudo potential for $\nu=1/3$ Laughlin state
\cite{Trugman-K} satisfies the conditions $V_{21}^2=V_{10}V_{30}$ and
$V_{20}>0$.  Then the Hamiltonian is rewritten as $\mathcal{H}_{1/3}=
\sum_{i} [Q^\dagger_i Q_i^{\mathstrut} + P^{\dagger}_i
P_i^{\mathstrut}], $ where $Q_i=\alpha_0 c_{i+1}c_{i+2}+\alpha_1
c_{i}c_{i+3}$ and $P_i=\beta_0 c_{i}c_{i+2}$ with $\alpha_0, \alpha_1,
\beta_0 \in {\mathbf R}$.  This Hamiltonian apparently has positive
expectation values $\braket{\mathcal{H}_{1/3}}\geq 0$, and the following
wave function as the exact ground state
\begin{equation}
\ket{\Psi}=\prod_i(1 + t c^\dagger_{i+1}
c^\dagger_{i+2} c_{i+3}^{\mathstrut}
c_{i}^{\mathstrut})\ket{\Psi_0},\quad
t \equiv -\alpha_1/\alpha_0,
\label{wf3}
 \end{equation}
since $Q_i\ket{\Psi_{1/3}}=P_i\ket{\Psi_{1/3}}=0, \forall i$ is
satisfied. This is three-fold degenerate but can be shown to be unique
ground state for each center-of-mass sectors.  

The wave function (\ref{wf3}) can be written in a MP form
$\ket{\Psi_{1/3}}= \mathrm{tr}[g_1 g_2 \cdots g_{N_e}]$
\cite{Fannes-N-W,Klumper-S-Z} where the matrix $g_i$ is identified in
the following way. First, we introduce the $S=1$ spin representation for
the three-sites unit cell as $\ket{010}\to\ket{\mathrm{o}}$,
$\ket{001}\to\ket{+}$ and $\ket{100}\to\ket{-}$.  One finds that the
possible configurations of the nearest two spins are $\ket{\mathrm{oo}}
+ t \ket{\mathrm{+}-} + \ket{\mathrm{o}+} +\ket{-\mathrm{o}}+ \ket{-+}$,
so that if we express the relations in the following $3 \times 3$ matrix
\begin{equation}
\label{f-matrix}
g_i=\left[
\begin{array}{c c c}
 f_{--}\ket{- }_i  &  f_{-\mathrm{o}}\ket{\mathrm{o} }_i
  &   f_{-+} \ket{+ }_i \\
 f_{\mathrm{o}-}\ket{- }_i  & f_{\mathrm{oo}}\ket{\mathrm{o} }_i 
  & f_{\mathrm{o}+} \ket{+ }_i \\
 f_{+-}\ket{- }_i  &  f_{+\mathrm{o}}\ket{\mathrm{o} }_i 
  &  f_{++}\ket{+ }_i
\end{array}
\right],
\end{equation}
its non-vanishing elements are identified as
$f_{\mathrm{oo}}=f_{\mathrm{o}+}= f_{-\mathrm{o}}= f_{-+}= 1$ and
$f_{+-}=t$. Next, we reduce the obtained $3\times 3$ matrix to a $2
\times 2$ form by changing the base from ($t\ket{-}, \ket{\mathrm{o}},
\ket{+}$) to ($t\ket{-}+\ket{\mathrm{o}}, \ket{+}$) as
\begin{equation}
g_{i}^{\mathstrut} \equiv
  \left[
   \begin{array}{cc}
    \ket{\mathrm{o}}_{i} & \ket{+}_{i} \\
    t\ket{-}_{i} & 0 
   \end{array}
\right].
\label{g-matrix3}
\end{equation}
The matrix (\ref{g-matrix3}) has a $4\times 4$ transfer matrix $G\equiv
\bar{g}_j \otimes g_j$ with only two eigenvalues $\lambda_{\pm}=(1\pm
\sqrt{4t^2+1})/2$. The argument for the Hamiltonian with an exact MP
wave function can also be extended to general $\nu=1/q$ Laughlin states
\cite{Wang-N}.

\section{Variational wave functions for excited states}

We derive excitation spectra at $\nu=1/3$ by a variational method.  In
the TT limit, the charge neutral excited states include one $01$-type
and one $0$-type domain walls that carry fractional charges $e^{*}=e/3$
and $e^{*}=-e/3$, respectively.  These states $\ket{\Psi_{0}^{\Delta
K=1}}=\ket{010 |0| 01 |010 |010\cdots}$, $\ket{\Psi_{0}^{\Delta
K=2}}=\ket{010 |0|010|01 |010\cdots}$, $\cdots$, are classified by the
center of mass coordinate relative to the ground state which is related
to the momentum of the $x$-direction.  Now we consider the variational
wave functions that consists of configurations generated from
$\ket{\Psi_{0}^{\Delta K}}$ by applying the Hamiltonian
$\mathcal{H}_{1/3}$. For $\Delta K=1$, there are following exchange
processes, $\ket{010| 001| 010| 010\cdots}\to$ $\ket{010| 001|
001|100\cdots}\to$ $\ket{010| 000| 110|100\cdots}\to$ $\ket{010| 001|
001|100\cdots}$.  Since the interactions in $\mathcal{H}_{1/3}$ are
limited within the third neighbors, there is no exchange process across
the defects.  We can rewrite these processes in the $S=1$ spin variables
by inserting vacancies appropriately (between the two $1$'s)
\cite{Nakamura-W-B1,Wang-T-N} as
\begin{equation}
\begin{split}
&\ket{\mathop{010}_{\mathrm{o}}|0| \mathop{010}_{\mathrm{o}}|
  \mathop{010}_{\mathrm{o}}| \mathop{010}_{\mathrm{o}}\cdots} \to 
  \ket{\mathop{010}_{\mathrm{o}}|0| \mathop{010}_{\mathrm{o}}| 
  \mathop{001}_{+}| \mathop{100}_{-}\cdots} \to \\
  &\ket{\mathop{010}_{\mathrm{o}}|0| \mathop{001}_{+}|
  \mathop{010}_{\mathrm{o}}| \mathop{100}_{-}\cdots}
 \to \ket{\mathop{010}_{\mathrm{o}}|0| \mathop{010}_{\mathrm{o}}|
 \mathop{001}_{+}| \mathop{100}_{-}\cdots}.
\end{split}
\end{equation}
We redefine the site indices so that the defect comes to the edge, and
introduce an ansatz with variational parameters $u$ and $v$ as $ \ket{
\Psi^{\Delta K=1}} =\mathrm{tr}[g_{1}' g_2^{\mathstrut} g_3^{\mathstrut} \cdots
g_{N_e-1}^{\mathstrut}g_{N_e}'']$, where
\begin{equation}
 g_{1}'\equiv
 \left[
    \begin{array}{c c}
     u\ket{\mathrm{o}}_{1} &  0 \\
     t\ket{-}_{1} & v\ket{\mathrm{o}}_{1}
    \end{array}
   \right], \quad
 g_{N_e}'' \equiv
 \left[
    \begin{array}{c c}
     \ket{\mathrm{o}}_{N_e} &0\\
     t\ket{-}_{N_e} & 0
    \end{array}
   \right].
\end{equation}
For the case $\Delta K \ge 2$, the ansatz should include three
variational parameters $u$, $v$ and $w$ as $\ket{\Psi^{\Delta K}}
=\mathrm{tr} [ g_1^{\mathstrut}g_2^{\mathstrut} \cdots
g_{K-1}^{\mathstrut} g_{K}''' 
g_{K+2}^{\mathstrut} \cdots
g_{N_e-1}^{\mathstrut} g_{N_e}'']$ with
\begin{equation}
 g_{i}'''\equiv
  \left[
   \begin{array}{cc}
    \ket{\mathrm{oo}}_{i,i+1} & u\ket{\mathrm{o}+}_{i,i+1}+v\ket{+\mathrm{o}}_{i,i+1} \\
    tu\ket{-\mathrm{o}}_{i,i+1}+tv\ket{\mathrm{o}-}_{i,i+1} & tw\ket{-+}_{i,i+1} 
   \end{array}
\right].
\end{equation}
\begin{figure}[ht]
\vspace{-0.8cm}
\begin{minipage}{0.5\hsize}
\centering
\includegraphics[width=8cm]{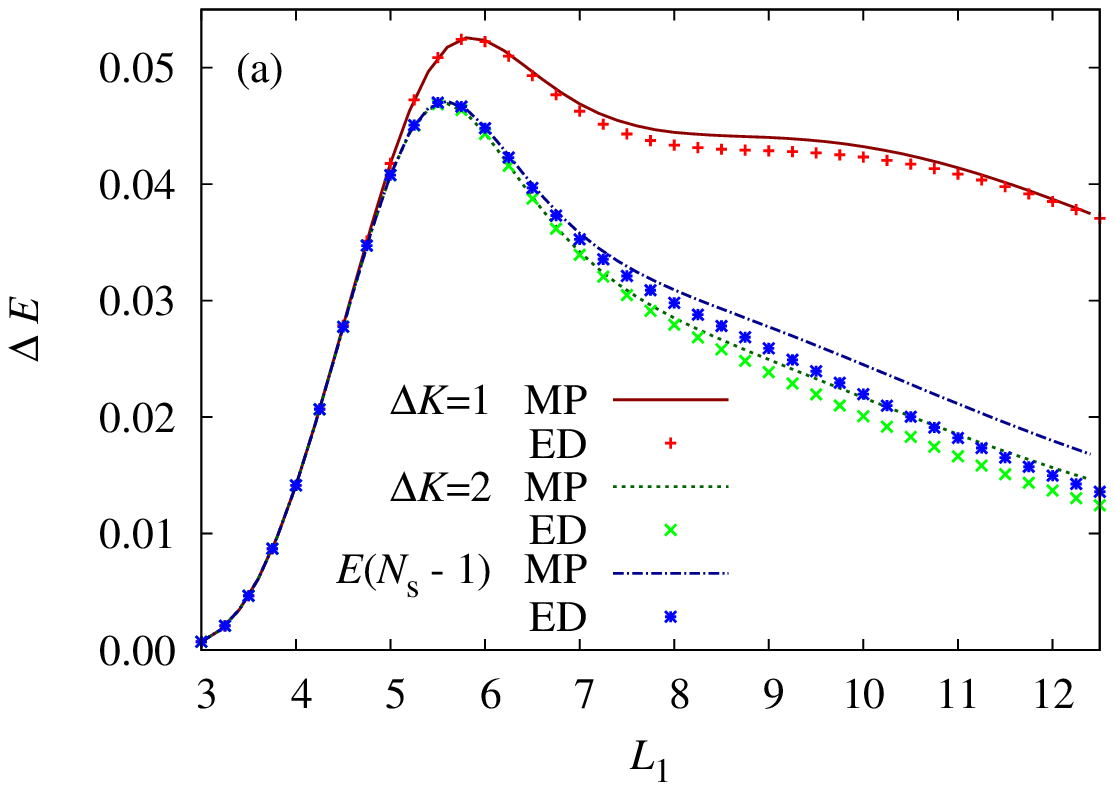}
\end{minipage}
\begin{minipage}{0.5\hsize}
\centering
\includegraphics[width=8cm]{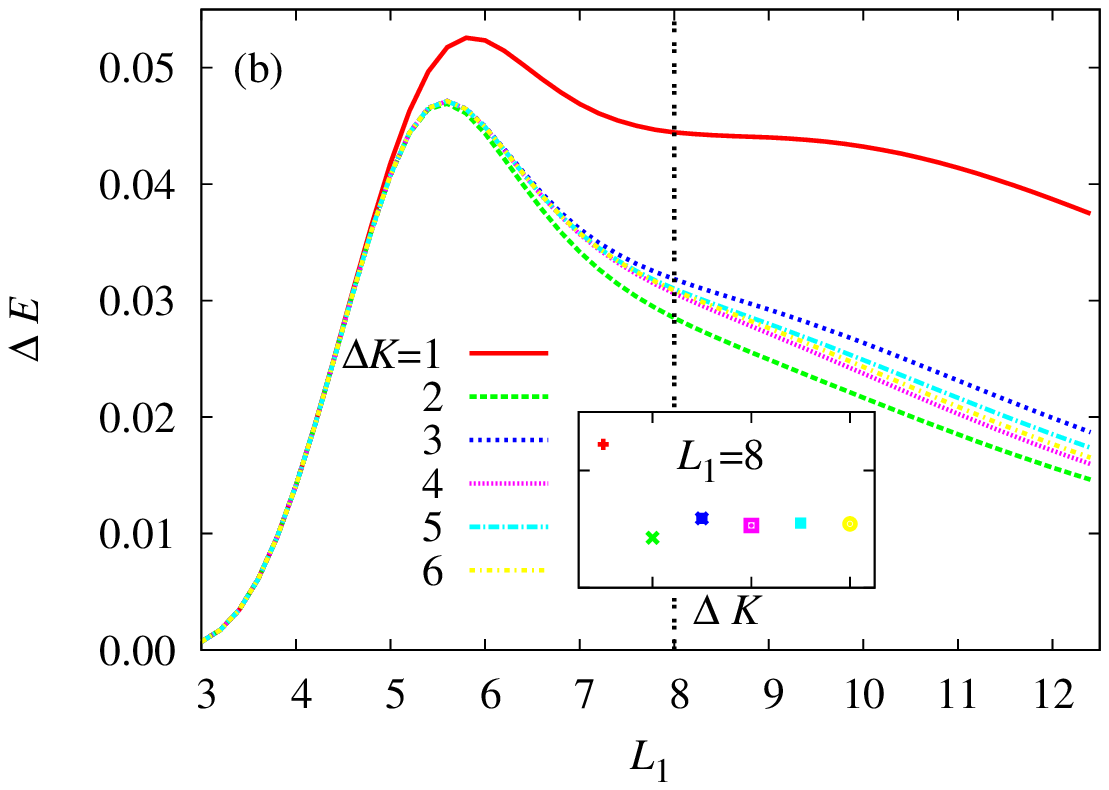}
\end{minipage}
\caption{(a) Comparison of excitation spectra obtained by the MP anzats
for infinite systems and those by exact diagonalization (ED) for a
finite system $N_s = 27$. (b) Neutral excitation spectra for different
momenta $\Delta K=1$-$6$ derived from the MP anzats for a infinite
system.  The inset shows the excitation spectra as function of $\Delta
K$ at $L_1=8$ which show a magneto-roton minimum.}  \label{ExSpec}
\end{figure}
When we calculate the compressibility, we need to shrink the system size
as $N_s \to N_s-1$ by removing 0 from the root state. Then only one
$01$-type domain wall ($\ket{\cdots010|010|01|010\cdots}$) should be
considered, so that the ansatz becomes $\ket{
\Psi^-}=\mathrm{tr}[g_{1}'''g_3^{\mathstrut} g_4^{\mathstrut} \cdots
g_{N_e-1}^{\mathstrut}]$.

These variational wave functions for $N_e \to \infty$ can be optimized
analytically by using the undetermined coefficient method.
Fig.~\ref{ExSpec}~(a) shows the excitation spectra obtained by the MP
ansatz for the infinite system as functions of $L_1$. These well agree
with the results obtained from the infinite systems by exact
diagonalization (ED). The deviation in the large $L_1$ region is due to
the finite size effect which appears through $\lambda_-/\lambda_+$.  We
also get a finite value of $E(N_s-1)$ for $N_e \to \infty$, reflecting
that the Laughlin state is incompressible
$\kappa^{-1}=\lim_{N_s\to\infty}(N_s/4\pi l_{\rm B}^2)
[E(N_s-1)+E(N_s+1)-2E_s(N_s)]\to\infty$ where $E(N_s+1)=E_s(N_s)=0$.
Fig.~\ref{ExSpec}~(b) shows the neutral excitation spectra obtained by
the MP ansatz for $\Delta K=1$-$6$.  The $\Delta K=2$ state is always
the lowest, and the excitation energy converges to a constant value when
$\Delta K$ is increased as shown in the inset.  This captures a
character of the magneto-roton behavior in the FQH state
\cite{Girvin-M-P}.  These results are also consistent with a recent
analysis based on spherical coordinate and the Jack polynomials
\cite{Yang-H-P-H}.

\section{Conclusion}
We have studied the excitation spectra of the $\nu=1/3$ FQH states
considering toroidal boundary conditions in the vicinity of the
Tao-Thouless limit. In this region, the original 2D system can be mapped
onto a 1D lattice model with short-range interactions which has the
exact MP ground state.  We have introduced ``ansatz'' for the excited
state as the MP state with defects and variational parameters, and
derived the excitation spectra analytically.  The excitation spectrum
with a fractional charge reflects the incompressibility of fractional
quantum Hall states.  The charge neutral excitations behave like the
roton mode.

\section{Acknowledgment}
We thank Emil J. Bergholtz for the related collaboration in
Ref.~\cite{Nakamura-W-B}.  Z.-Y.~W. and M.~N. acknowledge support from
the Global Center of Excellence Program ``Nanoscience and Quantum
Physics'' of the Tokyo Institute of Technology.  M.~N. also acknowledges
support from Grant-in-Aid No.23540362 by MEXT.



\begin{thebibliography}{99}
 \bibitem{Tao-T}
	 R.~Tao and D.~J.~Thouless:
	 Phys.\ Rev.\ B\ {\bf 28} (1983) 1142.

 \bibitem{Rezayi-H}
	 E.~H.~Rezayi and F.~D.~M.~Haldane:
	 Phys.\ Rev.\ B\ {\bf 50} (1994) 17199.

 \bibitem{Bergholtz-K}
	 E.~J.~Bergholtz and A.~Karlhede:
	 Phys.\ Rev.\ Lett.\ {\bf 94} (2005) 026802.

 \bibitem{Seidel-F-L-L-M}
	 A.~Seidel, H.~Fu, D.~-H.~Lee, J.~M.~Leinaas and J.~Moore:
	 Phys.\ Rev.\ Lett.\ {\bf 95} (2005) 266405.

 \bibitem{Laughlin}
	 R.~B.~Laughlin:
	 Phys.\ Rev.\ Lett.\ {\bf 50} (1983) 1395.

 \bibitem{Nakamura-W-B}
	 M. Nakamura, Z.~-Y.~Wang and E.~J.~Bergholtz: 
	 Phys.\ Rev.\ Lett.\ {\bf 109} (2012) 016401.

 \bibitem{Girvin-M-P}
	 S.~M.~Girvin, A.~H.~MacDonald and P.~M.~Platzman:
	 Phys.\ Rev.\ Lett.\ {\bf 54} (1985) 581;\\
	 Phys.\ Rev.\ B\ {\bf 33} (1986) 2481.

 \bibitem{Trugman-K}
	 S.~A.~Trugman and S.~A.~Kivelson:
	 Phys.\ Rev.\ B\ {\bf 31} (1985) 5280.

 \bibitem{Fannes-N-W}
	 M. Fannes,~B.~Nachtergale and R.~F.~Werner:
	 Europhys. Lett. {\bf 10} (1989) 633;\\
	 Commun. Math. Phys. {\bf 144} (1992) 443.

 \bibitem{Klumper-S-Z}
	 A. Kl\"{u}mper, A. Schadschneider and J. Zittartz:
 	 Z. Phys. B {\bf 87} (1992) 281;
	\\ Europhys. Lett. {\bf 24} (1993) 293.

 \bibitem{Wang-N}
	 Z.~-Y.~Wang and M.~Nakamura:
	 arXiv:1206.3071v2.

 \bibitem{Nakamura-W-B1}
	 M.~Nakamura, Z.~-Y.~Wang and E.~J.~Bergholtz: 
	 J. Phys.: Conf. Ser. {\bf 302}, 012020 (2011).

 \bibitem{Wang-T-N}
	 Z.~-Y.~Wang, S.~Takayoshi and M.~Nakamura:
	Phys. Rev. B {\bf 86} (2012) 155104.

 \bibitem{Yang-H-P-H}
	 B.~Yang, Z.-X. Hu, Z. Papic and F.~D.~M. Haldane:
	 Phys. Rev. Lett. {\bf 108} (2012) 256807.
\end{thebibliography}
\end{document}